# Investigation of Wireless Channel Asymmetry in Indoor Environments

Ehab Salahat[1], Ahmed Al-Tunaiji[1], Nazar Ali[1], and Raed M. Shubair[1,2]
[1]Etisalat British Telecom Center (EBTIC) & Khalifa University, UAE
[2]Research Laboratory of Electronics, Massachusetts Institute of Technology, USA

*Abstract*—Asymmetry is unquestionably an important characteristic of the wireless propagation channel, which needs to be accurately modeled for wireless and mobile communications, 5G networks, and associated applications such as indoor/outdoor localization. This paper reports on the potential causes of propagation asymmetry. Practical channel measurements at Khalifa University premises proved that *wireless channels* are *asymmetric* in realistic scenarios. Some important conclusions and recommendation are also summarized.

*Keywords—Channel Symmetry; channel asymmetry; indoor measurements; wireless channel; measurement setup.*

## I. INTRODUCTION

Signal propagation in wireless channels can be subjected to many parameters that degrade its performance. Such factors include noise, interference, large-scale fading (shadowing), small-scale fading, path loss, delay and other temporal and spatial dynamics that act as sources of impairments. As such, modeling radio channels is unquestionably a challenging endeavor and an accurate characterization is of a great benefit for the design of resilient future 5G wireless protocols and other applications.

A common assumption adopted in the wireless communications literature is channel symmetry. Some of the propagation impairments are assumed to have an equivalent effect when the roles of the transmitter and the receiver are interchanged (while maintaining the same link setup) resulting in an identical channel response. This also follows the well-known Friis's free-space transmission model:

$$P_r(d) = P_t \mathcal{G}_t \mathcal{G}_r \left[\frac{\lambda}{4\pi d}\right]^2 \quad (1)$$

where $P_t$, $\mathcal{G}_t$, $\mathcal{G}_r$, and $d$ respectively denote the transmitter power, transmitter gain, receiver gain, and the distance. This model is incorporated with the path loss model in any given measurement environment,

$$P_r(d,\beta) = P_r(d_0)\left(\frac{d_0}{d}\right)^\beta \quad (2)$$

where $\beta$ and $d_o$ denote the path loss exponent (e.g. $\beta=2$ in free-space) and the reference distance, respectively. Channel symmetry also coincides with the electromagnetic reciprocity theorem [1]. However, the theorem does not include the effect of impairments which cause channel asymmetry especially in presence of intermittent noise and interferences [2][3].

Radio asymmetry can have a big effect on many sensitive applications, e.g. frequency synchronization and location-based services in vehicular and mobile networks using time-of-arrival (ToA) and time-difference of arrival (TDoA), where both the forward and reverse channel responses affect the accuracy of these sensitive algorithms. To the best of the authors' knowledge, all literature on localization algorithms has assumed channel symmetry.

## II. CHANNEL ASYMMETRY – A SYNOPSIS

A summary of the potential causes of the asymmetric channel behavior is briefly discussed with an example.

### A. Symmetry Impairments

The term "asymmetry" is used to define the instantaneous characteristics of the forward and reverse channels of the propagating signal in a specific medium.

*1) Transmission Power*

Asymmetric transmitted power from, for example, a limited power mobile device to a base station can cause disparity in the received signal. Wireless networks are consist of many devices, and asymmetric links are very likely due to this form of such asymmetric transmitted powers [4].

*2) Hardware Sensitivity*

The hardware used in measurements can be a source of asymmetry. It was shown in [4] that devices manufactured by the same vendors but sold at different times differ in their transmission behavior. Hardware sensitivity was reported to be a major contributor to the asymmetry in [2].

*3) Antenna Design and Configuration*

Another source of link asymmetry is the antennas [4], especially with multiple antenna systems (i.e. spatial diversity such as in SIMO, MISO, and MIMO). The degree of asymmetry can depend on the directivity of the antenna and its configuration, and the utilized algorithms to estimate channel conditions. In addition, antenna's height and orientation are potential causes of channel asymmetry [5].

*4) Interference and Noise*

Interference and noise are likely to cause temporal and site-specific variations due their impulsive nature. Base stations transmitting signals from high heights will not experience interference similar to that of others transmitting in a congested and crowded city [3].

*5) Spectrum Shifting and Frequency Mismatch*

A study in [6] showed that "a large variation is manifested not only in the transmitted power, but also in the operational frequencies" of their WSN. This mismatch (misalignment) may be correlated with the hardware or the transmitted power.

*B. An Illustrative Example*

A mobile station (MS) such as a vehicle, can be located by triangulating three base stations (BS). Each BS estimates the location of the MS by estimating the total travel time from the power delay profile [7] obtained from the inverse Fourier transform of the channel frequency response. Since the object is moving, the frequency response of the forward path is different from the return path. If channel symmetry is assumed, then the travel time from the BS to MS will be the same, which is incorrect. Therefore, a more accurate position estimation can be obtained by considering the channel asymmetric.

### III. MEASUREMENT SETUP

A similar setup to the one described in [7-8] was used to conduct frequency channel measurements. The measurements were conducted at Khalifa University building in Sharjah, UAE using a Vector Network Analyzer (VNA) and two identical omnidirectional antennas positioned about 1.5 meters above the ground. Therefore, the directivity and antenna orientation are no longer a cause for concern. Moreover, since the VNA is calibrated, the issue of hardware sensitivity, transmission power or spectrum shifting should equally affect both paths.

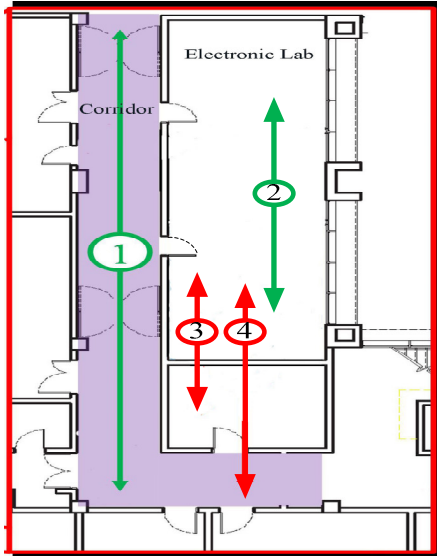

Fig. 1: Floorplan of Khalifa University, and measurement locations.

### IV. MEASUREMENTS: RESULTS AND ANALYSIS

Multiple indoor measurements at Khalifa University were conducted from which we report a few due to space limitations. As shown in Figure 1, LOS measurements, traces 1 and 2, were in the corridor and electronic lab (heavily cluttered), respectively while the NLOS measurements, traces 3 and 4, were between the lab, and the adjacent room (one wall separation) or the corridor, two walls separation. Exemplary results are shown in Figs. 2-3. Given that most of the known symmetry impairments were eliminated, both figures show symmetric forward and reverse channel behaviors even in multipath rich environment. This not only applies to the LOS, but also to the NLOS regardless of the partitioning medium and the distance separating the antennas. Although the measurements were in static conditions, human movements are unlikely to cause any asymmetry, as the speed of measurements are much faster than the speed of a human. Nevertheless, it is expected in practical environments, the wireless channel should be considered asymmetric because of the presence of impairments that may affect the signal in one path differently from the other.

### V. CONCLUSIONS

In this paper, measurements were used to prove that wireless channels can only be symmetric in the absence of practical temporal and spatial impairments. The asymmetric assumption should always be considered when modeling wireless channels in various environments. Measurements at Khalifa University campus have shown that multipath fading, transmitter-receiver separation, and LOS/NLOS propagation are not contributors to channel asymmetry.

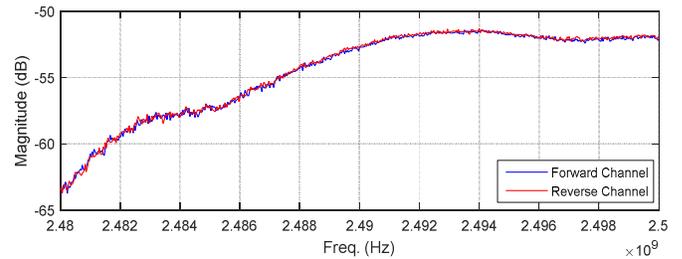

Fig. 2: Scenario 2 in Fig. 1, LOS measurements in the Lab.

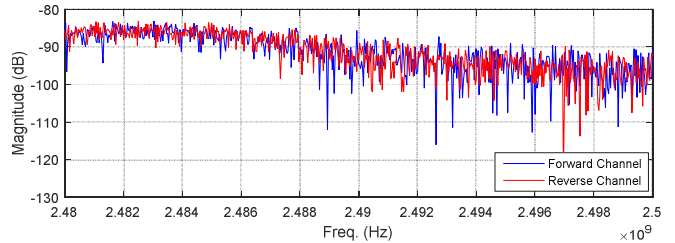

Fig. 3: Scenario 4 in Fig. 1, lab-to-corridor NLOS measurements.